\documentclass[superscriptaddress,onecolumn]{revtex4}
\usepackage{amsmath}
\usepackage{amsfonts}
\usepackage{adjustbox}
\usepackage{graphicx}
\usepackage{hyperref}
\usepackage{bbm}
\usepackage{doi}
\usepackage{xcolor}
\usepackage{multirow}

\begin{document}
%\title{Chiral Escape, Twist Solitons and Splitting of Defect Lines in Cholesteric Liquid Crystals}
\title{Escape into the Third Dimension in Cholesteric Liquid Crystals}
\author{Joseph Pollard}
\email{joe.pollard@unsw.edu.au}
\affiliation{School of Physics and EMBL Australia Node in Single Molecule Science, School of Medical Sciences, University of New South Wales, Sydney 2052, Australia.}  
\author{Gareth P. Alexander}
\email{G.P.Alexander@warwick.ac.uk}
\affiliation{Department of Physics, University of Warwick, Gibbet Hill Road, Coventry, CV4 7AL, United Kingdom.}

\begin{abstract}
  Integer winding disclinations are unstable in a nematic and are removed by an `escape into the third dimension', resulting in a non-singular texture. This process is frustrated in a cholesteric material due to the requirement of maintaining a uniform handedness and instead results in the formation of strings of point defects, as well as complex three-dimensional solitons such as heliknotons that consist of linked dislocations. We give a complete description of this frustration using methods of contact topology. Furthermore, we describe how this frustration can be exploited to stabilise regions of the material where the handedness differs from the preferred handedness. These `twist solitons' are stable in numerical simulation and are a new form of topological defect in cholesteric materials that have not previously been studied.  
\end{abstract}
\maketitle

\section{Introduction}

Escape into the third dimension is the colloquial name for a paper by Robert Meyer~\cite{meyer1973} in which he described how a liquid crystal confined to a cylindrical capillary could avoid a line singularity along its axis by allowing the molecules to tilt, up or down, so as to point along the cylinder axis, thereby `escaping into the third dimension'. The same effect was described contemporaneously by Cladis \& Kleman~\cite{cladis1972}. 
These escaped structures are commonly observed in experiments on materials confined in cylindrical capillaries~\cite{cladis1979,lequeux1988}, and they also occur in torodial droplets of liquid crystal~\cite{pairam2013,vitelli2014,fialho2017}, at the centre of a Skyrmion~\cite{machon2016PRSA,machon2016}, and as part of three-dimensional knotted solitons~\cite{tai2019,wu2022}. 
In this paper we consider the chiral version of Meyer's escape in which the director maintains a preferred handedness of twist. 

Chiral materials display an especially rich variety of topological states and offer many prospects for metamaterials, sensors and soft photonics~\cite{tai2019,wu2022,lin2011,lee2016,Mysliwiec2021}. In cholesterics, escaped structures are $\lambda$-lines, defects in the material's optical axis. They are important for photonic properties and lie at the heart of structures observed in many recent experiments~\cite{chen2013,tai2019,wu2022,omori2022,pieranski2023a, pieranski2023b}. 
In addition, many structures of biopolymers have been identified with cholesteric, or chiral, liquid crystal textures~\cite{bouligand1972,bouligand2008,ling2018,berent2022}, and in this vein the properties of cholesterics may find applications in tissue mechanics and developmental biology~\cite{whitfield2017,kole2021}. 

A significant advance in our understanding of topology in cholesterics has come from the introduction of methods of contact topology~\cite{geiges2008,machon2017}. These have been used to demonstrate the preservation of the layer structure in a cholesteric~\cite{machon2017}; to explain the stability of Skyrmions in liquid crystals and chiral magnets~\cite{hu2021}; to analyse defect structures in cylindrical capillaries~\cite{eun2021}; to shed light on the transition pathways between different cholesteric textures~\cite{han2022}; and to extend the usual classification of point and line defects in an achiral material to account for chirality~\cite{pollard2019, pollard2023}. Contact topological methods thus have broad applicability to the study of cholesterics and chiral materials more generally, and their development has utility beyond any given application. 

The properties we describe here concern the basic conditions of compatibility of chirality with three-dimensional variation of the director and are relevant to all aspects of cholesterics, from metamaterials to biological settings. 
Via a novel application of contact topology, we show that escape in a cholesteric may be frustrated by the requirement that the director maintain a uniform sense of handedness. Frustration leads to a wider array of metastable structures than can be observed in an achiral system, including stable strings of point defects in spherical droplets~\cite{sec2012}, shells~\cite{darmon2016} and cylindrical capillaries~\cite{eun2021}, as well the knotted and twisted structures that appear in spherical droplets~\cite{robinson1958,sec2012}, at the core of heliknotons and Hopfions~\cite{tai2019,wu2022}, and have been observed forming from a Lehmann cluster in recent experiments~\cite{omori2022,pieranski2023a,pieranski2023b}. Our methodology offers a new perspective on these structures and provides analytic models for describing the processes that underlie their formation, as well as defect splitting processes more generally. 

We show further that frustration results in a new type of topological defect in a cholesterics, the twist soliton. These consist of regions of the material in which the handedness is opposite from the energetically preferred handedness. Such defects have been observed previously~\cite{posnjak2017,pollard2019,ackerman2016}, but we do not know of a systematic study. We describe the conditions that give rise to these defects and give experimentally-accessible examples of settings in which they are protected by topological constraints. 

The remainder of this paper is organised as follows. In Section \ref{sec:achiral_escape} we recall the scenario studied by Meyer~\cite{meyer1973} of an achiral nematic confined in a cylindircal capillary, along with the homotopy theory treatment of this structure. An achiral director, singular along the capillary, escapes either up or down with equal probability, while in a cholesteric the direction of escape sets the handedness when the boundary anchoring is tangential. In Section \ref{sec:local_obstructions} we introduce techniques from contact topology to explain this phenomenon and describe more generally the conditions that force a reversal of the director's handedness---readers who are familiar with contact topology or only interested in the applications may skip this section. This theory is applied in Sections \ref{sec:tau1}, \ref{sec:chi1}, and \ref{sec:chi2} to describe the frustration that occurs when removing $\chi$-lines in a cholesteric and to explain the resulting textures. We also describe how to force regions of reversed handedness to occur and show these are metastable in simulations. The paper then concludes in Section \ref{sec:discussion} with a discussion.

\section{Escape in a Cylinder} 
\label{sec:achiral_escape}

We begin by reviewing Meyer's theory of escape into the third dimension~\cite{meyer1973,cladis1972} and the related homotopy theory description of escaped textures. Consider a singular line in a nematic director. The winding of the singular line may be measured by taking a generic disc $D$ orthogonal to the singularity and computing the winding number of the director around the boundary of the disc. For a stable defect (disclination) this winding is fractional, generically $\pm 1/2$, and may vary along the singular line. In a coreless defect the winding is an integer, generically $\pm 1$. Depending on confinement and material properties, a $\pm 1$-winding singular line in an achiral nematic may either split into a pair of $\pm 1/2$ disclination lines or else be removed by escape~\cite{meyer1973,cladis1979,lequeux1988,murray2017,velez2021}.

Meyer considered the case of an achiral nematic in a cylindrical capillary of radius $R$ with radial boundary conditions. A natural texture has the director everywhere radial with a singularity of winding $+1$ along the axis, i.e., ${\bf n} = {\bf e}_r$ in cylindrical coordinates $r,\theta, z$. Meyer showed that this structure is always unstable (provided $R$ is larger than the core size), as the energy can be reduced by removing the singularity via an `escape' described by the director 
\begin{equation}
    {\bf n} = \cos\alpha \,{\bf e}_{z} + \sin\alpha \,{\bf e}_{r},
\end{equation}
in terms of an angle $\alpha$ satisfying $\alpha=\pi/2$ when $r=R$ and $\alpha = 0,\pi$ along the cylinder axis $r=0$. We consider the Frank free energy with a one-elastic-constant approximation and look for minimisers with $\alpha = \alpha(r)$ a function of the radial coordinate only. These are well-known to give rise to the Bogomol'nyi-Prasad-Sommerfield (BPS) form 
\begin{equation}
    F = \int \frac{K}{2} \biggl( \partial_r \alpha \mp \frac{\sin\alpha}{r} \biggr)^2 dV + 2\pi K L , 
\end{equation}
where $L$ is the length of the cylinder and the $\pm$ sign relates to the two possible directions for the escape. The BPS bound, $2\pi K$ per unit length, is attained for 
\begin{equation}
    \alpha = \begin{cases} 2 \arctan \frac{r}{R} & \textrm{escape up} , \\ \pi - 2 \arctan \frac{r}{R} & \textrm{escape down} . \end{cases}
    \label{eq:BPS_escape_theta}
\end{equation}
These states are distinct in that they represent different homotopy classes of escape profile. The director on a cross-sectional disc of the cylinder corresponds to a map of the disc into $\mathbb{RP}^2$ that is constrained on the boundary. For the topological classification we can take the boundary condition to be that the director lies in an `equatorial' $\mathbb{RP}^1$ of $\mathbb{RP}^2$, which gives a classification in terms of the relative homotopy group $\pi_2(\mathbb{RP}^2,\mathbb{RP}^1) \cong \mathbb{Z}^2$~\cite{machon2016PRSA}. Concretely, the homotopy type of the director is a pair of integers $(p,q)$ which count the number of times the director wraps around the northern and southern hemispheres of a sphere, with the difference $p-q$ being the winding of the director around the boundary of the cylinder. In the present case the boundary condition imposes $p-q=+1$. The `escape up' and `escape down' textures then correspond to the elements $(+1,0)$ and $(0,-1)$, respectively. 

\begin{figure*}[t]
 \centering
 \includegraphics[width=\linewidth]{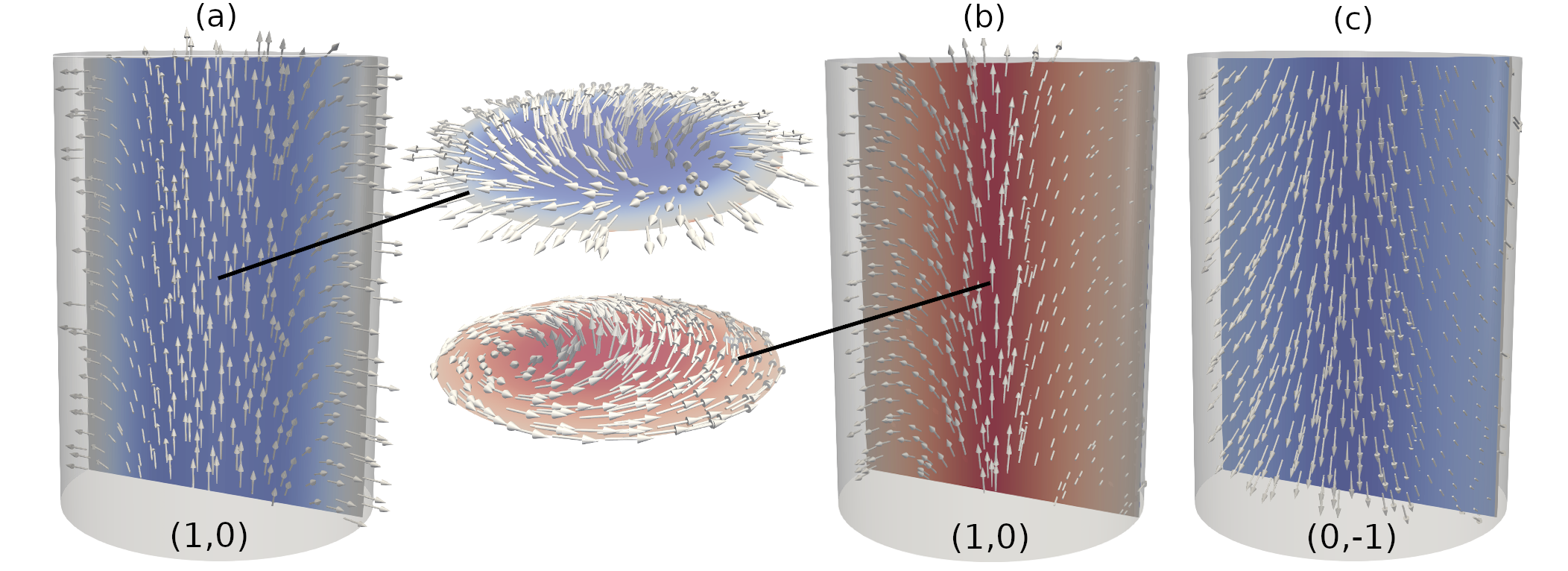}
 \caption{Escape of a $+1$ singularity in a cylinder. The twist is shown on slices, with blue being negative (right-handed) and red positive (left-handed), and the oriented director is shown as white arrows. (a) When there are radial boundary conditions, the director may escape either up or down, and in either case it is possible to make the director chiral, with the twist vanishing only on the boundary. We also show the director on a cross-sectional disc.  For tangential boundary conditions, there is difference between (b) escape up and (c) escape down, namely, the former results in a left-handed director and the latter a right-handed one. In order to better illustrate the direction of escape, we have oriented the director so that it points outwards from the boundary in (a), and so that it is flowing anticlockwise around the boundary in both (b) and (c). In each case the panel is labelled by the element $(p,q) \in \pi_2(\mathbb{RP}^2,\mathbb{RP}^1)$ the texture corresponds to.}
 \label{fig:escape}
\end{figure*}

Both cases are achiral, but can be made chiral by a small perturbation. Specifically, writing ${\bf n} = \cos\beta (\cos\alpha \,{\bf e}_{z} + \sin\alpha \,{\bf e}_{r}) + \sin\beta \,{\bf e}_{\theta}$ we find
\begin{equation}
    {\bf n} \cdot \nabla \times {\bf n} = \sin\beta \cos\beta \,\sin\alpha \,\partial_r \alpha + \biggl( \partial_r \beta + \frac{\sin\beta \cos\beta}{r} \biggr) \cos\alpha .
\end{equation}
Linearising, we see that a choice such as $\beta \sim \mp J_1(kr)$, where $k$ is such that $kR$ is the first zero of the Bessel function $J_1$, gives a chiral texture at all interior points of the cylinder. Different signs need to be chosen depending on the desired handedness and whether it is escape up or escape down. We illustrate escape up in Fig.~\ref{fig:escape}(a), showing the twist on a slice across the cylinder, with red indicating positive and blue negative; the case of escape down is completely analogous. The director is chiral throughout, with the twist only vanishing on the boundary, as is required by the boundary condition. 

Meyer also considered the case of tangential boundary conditions, for which the equivalent singular structure is described by ${\bf n} = {\bf e}_\theta$. The escaped form of this structure is ${\bf n} = \cos\alpha \,{\bf e}_z + \sin\alpha \,{\bf e}_{\theta}$ and in a one-elastic-constant approximation the minimiser is again given by the BPS state~\eqref{eq:BPS_escape_theta}. This time the director is already chiral, with escape up being left-handed and escape down being right-handed. This correlation between direction of escape and handedness is the most basic form of the compatibility that we describe here. The two cases are shown in Figs.~\ref{fig:escape}(b,c) respectively, and we emphasise the distinction by orienting the director so that, in each case, it flows anticlockwise around the boundary. The twist is shown on a slice, coloured blue or red according to whether it is negative or positive. 

It is possible to have a right-handed texture that escapes down at the centre of the cylinder with tangential anchoring. We can achieve this by introducing an extra $\pi$ twist into the director as we move from the centre to the boundary of the disc. However, this texture is still represented by $(-1,0) \in \pi_2(\mathbb{RP}^2,\mathbb{RP}^1)$. To see this, note that the extra region of $\pi$ twist we have added is an annulus. As we move outwards from the inner boundary to the centre of the annulus, the director is mapped onto the entire southern hemisphere of $S^2$. As we continue on to the other boundary of the annulus we cover the southern hemisphere again, but we do so with the opposite orientation---thus, this annulus region we have added corresponds to the trivial class $(0,0)$.

A radial $+1$ singularity may decompose into a pair of $+1/2$ disclination lines instead of escaping. This process is naturally associated with chiral materials~\cite{cladis1979,lequeux1988}, but can also arise as a result of anisotropy in the elastic constants, for example in achiral materials which have a relatively small twist elastic constant compared to the bend and splay elastic constants~\cite{jeong2015,velez2021}. We revisit this situation in  \S\ref{sec:chi2} below. 

\subsection{Numerical Simulation}

In the examples shown in Fig.~\ref{fig:escape}, and throughout the paper, the director configurations are obtained from numerical simulations minimising the Frank free energy 
\begin{equation}
    F = \frac{K}{2} \int_M \bigl| \nabla {\bf n} \bigr|^2 + 2q_0 \,{\bf n} \cdot \nabla \times {\bf n} \,dV ,
    \label{eq:Frank_energy}
\end{equation}
where $M$ is the material domain, $K$ is a Frank elastic constant and the chirality $q_0$ is a parameter favouring non-zero twist and setting the cholesteric pitch, $2\pi/|q_0|$, with positive values giving right-handed twist. We adopt the one-elastic-constant approximation for simplicity. As our focus is not on disclination lines the use of director-based simulations is appropriate---they capture the phenomenological features that we wish to describe. Numerical minimisation is performed using a finite difference method on a cubic grid starting from initial conditions that reflect our analytical treatment of the structure in question. 
Depending on the structure being studied the material domain is a cylindrical capillary of length $R$, a spherical droplet of radius $R$, or a cube of side length $R$. At the boundary we impose either homeotropic or planar degenerate anchoring, depending on the structure being studied, and in our simulations we implement this as a hard constraint, corresponding to an infinite energy penalty for deviating from the desired anchoring. 

A close similarity exists between the textures of cholesteric liquid crystals and those of chiral ferromagnets~\cite{fukuda2011,leonov2014,machon2016}; indeed~\eqref{eq:Frank_energy} is exactly the energy of a ferromagnet with the Dzyaloshinskii-Moriya interaction. Thus, our results in this paper may also be applied to chiral ferromagnets. Note that ferromagnetics have a polar symmetry: while disclination lines cannot exist in these materials, the escaped defects we describe here can. Superfluid phases of $^3\text{He-A}$ are also chiral and also described by a vector order parameter~\cite{vollhardt1990}, and so the topological aspects of our results apply to these phases as well.

\section{Contact Topology and Chiral Directors Near Surfaces}
\label{sec:local_obstructions}

We now introduce methods from the mathematical field of contact topology that characterise the frustration that can arise when attempting to escape a singular line in a chiral material. We emphasise that these methods are part of a general framework for the analysis of chiral media, and therefore have broad applicability beyond the present application. 

Contact topology~\cite{geiges2008} involves the study of plane fields, rank 2 sub-bundles of the tangent bundle---that is, a choice of a plane in the tangent space at each point, with the planes varying smoothly over space. Any orientable plane field can be defined as the kernel of a differential 1-form $\eta$, with the planes being spanned by vector fields ${\bf v}$ satisfying $\eta({\bf v})=0$. A plane field where the planes rotate with a constant sense of handedness is called a contact structure, a condition which is encoded in the constraint $\eta \wedge d\eta \neq 0$. A 1-form satisfying this condition is called a contact form. If a director ${\bf n}$ is chiral, satisfying ${\bf n} \cdot \nabla \times {\bf n} \neq 0$, then the 1-form $\eta$ dual to the director---defined by $\eta({\bf v})={\bf n} \cdot {\bf v}$ for any vector field ${\bf v}$---is a contact form and the associated contact structure consists of those planes that are orthogonal to ${\bf n}$. This observation establishes a duality between chiral directors and contact structures which allows us to use results from contact topology to study cholesterics~\cite{machon2017,pollard2019,eun2021,han2022,pollard2023}.

\subsection{Surfaces and Characteristic Foliations}
\label{sec:local_obstructions1}

When considering geometric structures on manifolds it is instructive to examine their behaviour close to submanifolds of lower dimension---in three dimensions these are points, curves, and surfaces---which often have associated topological and geometric invariants that provide insight into the full structure. Many results in contact topology come from this approach, which we briefly summarise. All contact structures are equivalent in a neighbourhood of a point---this is the Darboux theorem~\cite{geiges2008}. There are two classes of curves naturally associated to a contact structures, those that are everywhere transverse to the contact planes and those that are everywhere tangent. The latter are called Legendrian curves, and they have an associated invariant called the Thurston--Bennequin number. The physical meaning of this invariant has been discussed in previous work~\cite{machon2017,pollard2023}: it provides a count of cholesteric layers. In this work we consider the local structure of a chiral director close to a surface $S$ with normal vector ${\bf N}$. If $S$ has boundary, we assume it is either a Legendrian curve, or else parallel to the director. Generically, we may assume that the director is orthogonal to the interior of the surface only at isolated points and tangent to it along a set of disjoint closed curves---in the contact topology literature such surfaces are called convex~\cite{geiges2008, pollard2023}. Surfaces inside a liquid crystal material can be imaged using optical microscopy~\cite{smalyukh2001,posnjak2016,posnjak2017,posnjak2018experimental}, and the projection of the director into the surface can be extracted from optical data---this means that the quantities we describe in this section can be directly measured in experiments.

\begin{figure*}[t]
 \centering
 \includegraphics[width=0.96\linewidth]{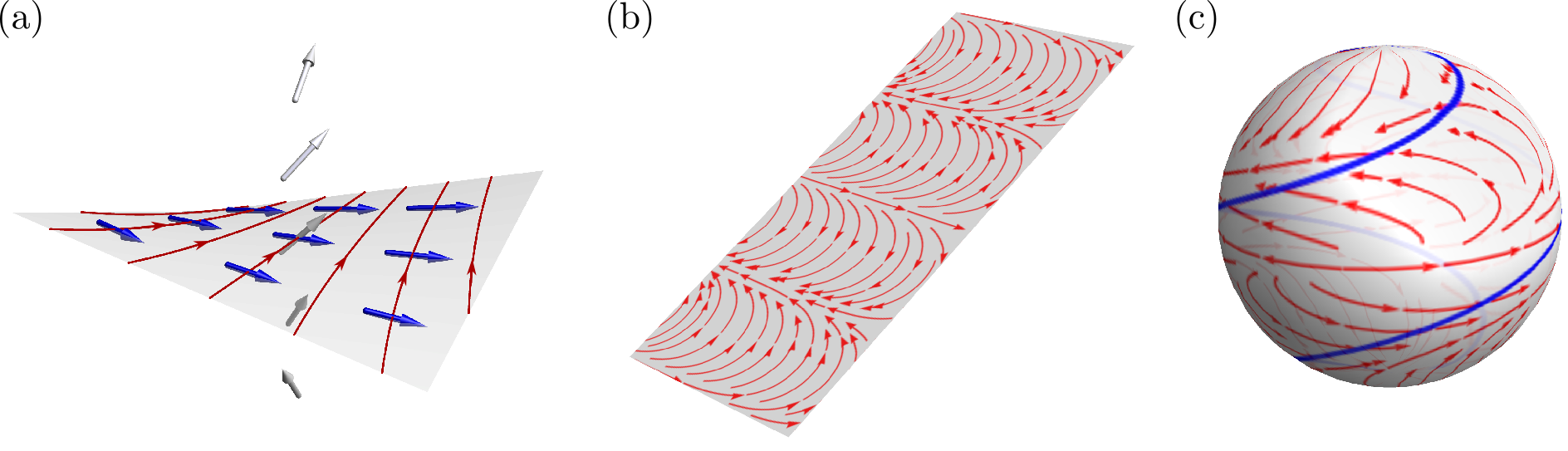}
 \caption{(a) General schematic of the characteristic foliation on a surface (light grey). The director is shown on a vertical line passing through the surface (grey arrows) and its projection into the surface, ${\bf n}_{\perp}$, is shown by the blue arrows. The red lines show the characteristic foliation, directed by ${\bf f} = {\bf N} \times {\bf n}_{\perp}$. (b) Example of the characteristic foliation on a plane surface for the helical cholesteric ground state with pitch axis along the vertical direction. (c) Same but for a spherical surface.}
 \label{fig:char_fol}
\end{figure*}

The intersection of the contact planes with a convex surface inscribes a `characteristic foliation' on the surface. In terms of the director, which we may write as ${\bf n} = {\bf n}_\perp + ({\bf n} \cdot {\bf N}) \ {\bf N}$ along the surface, it is more natural to consider the projection ${\bf n}_\perp$, which is orthogonal to the characteristic foliation but contains the same information. We will denote by ${\bf f}$ the vector field ${\bf N} \times {\bf n}_{\perp}$ that `directs' the characteristic foliation. In Fig.~\ref{fig:char_fol} we show this structure, giving a general schematic and two specific examples for convex surfaces for a director corresponding to the cholesteric ground state. 

As long as $S$ does not intersect any defects in the director field, then both the projection ${\bf n}_\perp$ and the characteristic foliation ${\bf f}$ on $S$ have singularities exactly at points where the director is orthogonal to $S$. Generically these are isolated points. If $S$ does intersect defects in ${\bf n}$ then (generically) it also does so at isolated points and ${\bf f}$ will have additional singularities at these points---we return to this case in \S\ref{subsec:local_obstructions2} and for now consider only the case where $S$ does not intersect defects. A singularity is therefore either an `escape up' singularity if the director points out of the surface there, or an `escape down' singularity if the director points into the surface. We will also describe this by an orientation $s=1$ for escape up and $s=-1$ for escape down. 

We give a local model for a zero in the characteristic foliation. Introduce a local coordinate system $x,y,z$ in a neighbourhood of a singular point $p$ of $S$ with the $z$-axis corresponding to the surface normal. In this neighbourhood of $p$ we can write ${\bf n} = n_x \,{\bf e}_x + n_y \,{\bf e}_y + s{\bf e}_z$, where $s=\pm 1$ is the orientation of the director at $p$. It is convenient to make use of complex coordinates $\zeta = x+iy$ on $S$ and to write $n_{\perp} = n_x + i n_y$. For a generic surface the zeros of this function have non-trivial linear part and at lowest order 
\begin{equation} \label{eq:char_fol_chiral1}
    n_{\perp} = c_{1} \zeta + c_{-1} \overline{\zeta} + \cdots .
\end{equation}
The winding number of the zero is $w=+1$ if $|c_{1}| > |c_{-1}|$ and $w=-1$ if $|c_{-1}| > |c_{1}|$. A direct calculation then gives 
% $\partial_x n_y - \partial_y n_x = 2 \,\mathrm{Im}\, \partial n_{\perp} = 2 \,\mathrm{Im}\, c_{1} + \cdots$, and consequently 
\begin{equation}  \label{eq:char_fol_chiral2}
    \Bigl. {\bf n} \cdot \nabla \times {\bf n} \Bigr|_{p} = 2 s \,\mathrm{Im}\, c_1 = -s \, \Bigl. \textrm{div}\, {\bf f} \Bigr|_{p} .
\end{equation}
The last equality relating the twist to the divergence of the characteristic foliation is the usual presentation of this result in contact topology~\cite{geiges2008}. We conclude that a director may be chiral close to a singularity $p$ in the characteristic foliation only when $\mathrm{Im}\,c_1$ is non-zero, and the handedness is correlated with the winding around the singularity and whether it is escape up or escape down, as summarised in Table~\ref{tab:escape}. This calculation explains the disparity, noted in the previous section, between escape up and escape down for tangential boundary conditions in a cylinder. For a chiral director then, singularities in the characteristic foliation are described by the local model $n_{\perp} = c_1 \zeta + c_{-1} \overline{\zeta}$ with $\mathrm{Im}\,c_1$ non-zero and its sign correlated with the sense of twist and whether it is escape up or escape down, as in Table~\ref{tab:escape}. 

\begin{table}[b]
    \begin{center}
        \begin{tabular}{c|c|c}
            & \textbf{Escape Up} & \textbf{Escape Down}\\
            \hline
            \multirow{2}{*}{\textbf{Right-handed}} & $\mathrm{Im}\, c_1 < 0$ & $\mathrm{Im}\, c_1 > 0$\\ 
            & $\textrm{div}\, {\bf f} > 0$ & $\textrm{div}\, {\bf f} < 0$\\
            \hline
            \multirow{2}{*}{\textbf{Left-handed}} & $\mathrm{Im}\, c_1 > 0$ & $\mathrm{Im}\, c_1 < 0$\\
            & $\textrm{div}\, {\bf f} < 0$ & $\textrm{div}\, {\bf f} > 0$\\
        \end{tabular}
        \caption{The relationship between the local structure of a zero in the characteristic foliation, the direction of escape and the handedness of twist.}
        \label{tab:escape}
    \end{center}
\end{table}

This is the only constraint on the characteristic foliation of a chiral director: as long as it is satisfied, it is possible to construct a chiral director in a neighbourhood of the surface imparting that characteristic foliation~\cite{geiges2008}. It is also generic: given any characteristic foliation on $S$, an arbitrarily small perturbation will make it satisfy this condition~\cite{arnold}. However, the constraint is not generic in a one-parameter family of characteristic foliations, which is the origin of the frustration we will describe presently. 

Besides the connection with chirality, there is additional topological information encoded in the singularities of the characteristic foliation. From the homotopy theory perspective each singularity corresponds to a meron~\cite{machon2016PRSA}, representing either a generator of the group $\pi_2(\mathbb{RP}^2,\mathbb{RP}^1)$ or its inverse. We may associate to each singularity a winding number $w = \pm 1$ and an orientation $s = \pm 1$. If $s=+1$ then the meron is escape up and corresponds to the group element is $(w,0) \in \pi_2(\mathbb{RP}^2,\mathbb{RP}^1)$, and conversely if $s=-1$ the meron is escape down and the corresponding group element is $(0,-w)$. If the director lies in $\mathbb{RP}^1$ along the boundary of $S$ then summing these group elements over all singularities yields the group element $(p,q) \in \pi_2(\mathbb{RP}^2,\mathbb{RP}^1) $ which represents the homotopy class of the entire texture. The difference $p-q$ is the sum of the windings of all singularities in the characteristic foliation, which by the Poincar\'e--Hopf theorem is equal to $\chi_E(S)$. 

Plane fields possess a quasi-2D topological invariant called the Euler class~\cite{bottandtu}, which can be measured across any surface $S$. Physically the Euler class corresponds to the Skyrmion charge $Q$ on $S$, and it can be computed via a signed count of the singularities in the characteristic foliation:
\begin{equation} \label{eq:euler_class}
  2Q = \sum_j s_j w_j,
\end{equation}
where $s_j, w_j = \pm 1$ are respectively the orientation and winding number of the $j$th singularity. We note that~\eqref{eq:euler_class} is more general than the usual integral formula for the Skyrmion charge as it can be applied to any surface $S$, whereas the integral formula is properly defined only when $S$ is closed---in that case, the two formulae agree. 

If the director is oriented, then the curves $\Gamma$ where ${\bf n} \cdot {\bf N} = 0$ divide the surface into two regions $S^+, S^-$ where the director points out of and into the surface respectively---$\Gamma$ is called the `dividing curve'~\cite{geiges2008,pollard2023}. By the Poincar\'e--Hopf theorem, the sum of the windings of the singularities is equal to the Euler class $\chi_E(S)$ of the surface. By the same theorem, the sum of the winding numbers of just the escape up singularities computes $\chi_E(S^+)$, and similarly the sum of the windings of the escape down singularities computes $\chi_E(S^-)$. Thus~\eqref{eq:euler_class} reduces to~\cite{geiges2008,pollard2023}
\begin{equation} \label{eq:euler_class2}
  2Q = \chi_E(S^+) - \chi_E(S^-).
\end{equation}
If $S$ is a sphere then $Q$ determines the defect charge within that sphere, and in this context the formula~\eqref{eq:euler_class2} has been given previously~\cite{copar2012}. We see that a basic meron has Skyrmion charge $\pm 1/2$; a meron is indeed a fractionalisation of a Skyrmion.

\begin{figure*}[t]
 \centering
 \includegraphics[width=\linewidth]{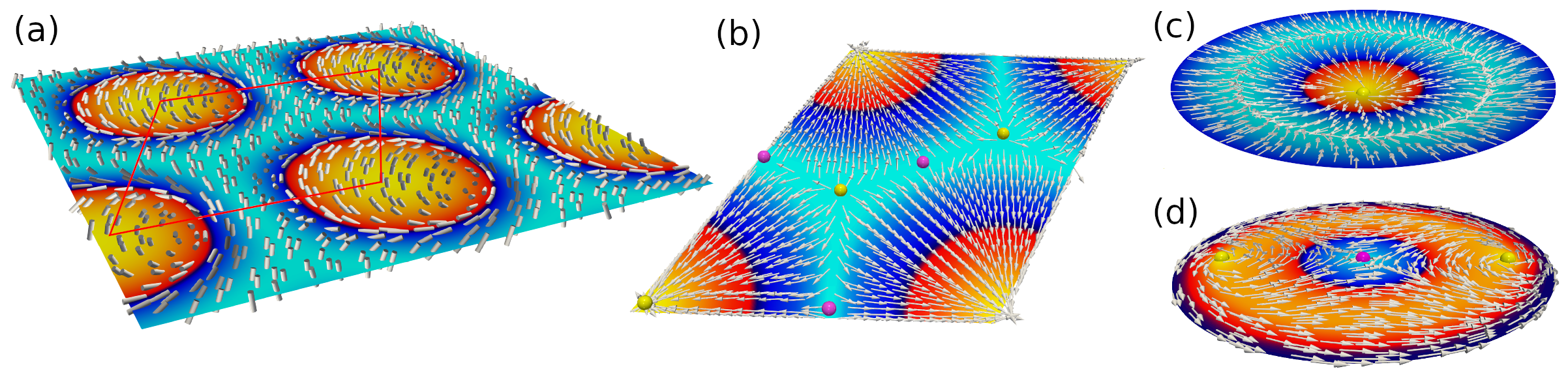}
 \caption{(a) A hexagonal Skyrmion lattice. The director (white sticks) is shown on a slice, which is coloured according to whether the director points out of (yellow) or into (blue) the surface. A single cell of the lattice is indicated in red. (b) A closeup of the parallelogram shown in panel (a)---left and right should be identified, as should top and bottom. The characteristic foliation ${\bf f} = {\bf e}_z \times {\bf n}$ is shown as white arrows. Its singularities are highlighted by spheres coloured yellow and purple according to whether the winding $w$ is $+1$ or $-1$ respectively. These singularities allow us to count the Skyrmion charge as described in the text. (c) The characteristic foliation on a cross-sectional disc of an escape texture in a cylindrical capillary with tangential anchoring. The single singularity with $s=1, w=1$ illustrates this texture has homotopy class $(1,0) \in \pi_2(\mathbb{RP}^2,\mathbb{RP}^1)$. (d) A more complex texture in a cylindrical capillary, this time with normal anchoring on the boundary. The singularities of the characteristic foliation show this texture has homotopy class $(2,1)$, as described in the text.}
 \label{fig:euler_class}
\end{figure*}

To illustrate these topological properties we take as an example the lattice of Skyrmions shown in Fig.~\ref{fig:euler_class}. The plane shown is a convex surface and is coloured yellow/blue according to whether the director points into/out of the surface. Each hexagonal cell of the lattice contains a central Skyrmion surrounded by six $\lambda^{-1/2}$-lines at the points of the hexagon. The lattice is made of tessellating parallelogram cells, as indicated by the red lines in Fig.~\ref{fig:euler_class}(a), each of which has Skyrmion charge 1. One such cell is shown in Fig.~\ref{fig:euler_class}(b). Accounting for the periodic boundary, the characteristic foliation on this unit cell (white arrows) contains one singularity with $s=+1, w=+1$, three singularities with $s=-1, w=-1$, and two with $s=-1, w=+1$---these singularities are indicated by spheres, coloured yellow if $w=+1$ and purple if $w=-1$. Putting these numbers into the formula~\eqref{eq:euler_class} yields a Skyrmion count of $Q=1$ for the unit cell, as expected. One may similarly note that $S^+$ is a disc, with $\chi_E(S^+)=1$, and $S^-$ is a torus with a disc removed, $\chi_E(S^-)=-1$, so~\eqref{eq:euler_class2} gives the same result.

We further illustrate this computation for an escape-up texture in a cylindrical capillary of radius $R$ with tangential boundary conditions, Fig.~\ref{fig:euler_class}(c). As described in Section \ref{sec:achiral_escape} the only way we can have a right-handed texture that escapes up at the core of capillary is if we add an additional $\pi$-twist to the director, ${\bf n} = \cos(3\pi r/2R){\bf e}_z - \sin(3\pi r/2R){\bf e}_\theta$. The characteristic foliation induced on a disc has a singulatity with $s=1, w=1$ at the origin, and an entire line of singularities at $r = 2R/3$. In order to perform the calculation we perturb the disc slightly so as to make the characteristic foliation generic, which replaces the line of singularities with a closed orbit. Thus, the characteristic foliation has a single singularity with $s=1, w=1$, which tells us this director corresponds to $(1,0) \in \pi_2(\mathbb{RP}^2,\mathbb{RP}^1)$. In Fig.~\ref{fig:euler_class}(d) we show for comparison a texture in a cylindrical capillary with tangential anchoring which corresponds to the element $(2,1) \in \pi_2(\mathbb{RP}^2,\mathbb{RP}^1)$. The homotopy type is readily seen from counting the singularities, of which there is one with $s=-1,w=-1$ (a meron whose homotopy type is $(0,1)$) and two with $s=1,w=1$ (merons whose homotopy type is $(1,0)$). 

Finally, the characteristic foliation relates to another topological invariant that is unique to chiral directors. A contact structure is called overtwisted when it contains an `overtwisted disc', which is nothing more than a $+1$-winding meron~\cite{geiges2008,machon2017,pollard2023}: the classical picture of an overtwisted disc is exactly the cross section of Fig.~\ref{fig:escape}(b). A director that is not overtwisted is called tight, and this is a dichotomy: representatives of the two classes cannot be homotoped into one another while keeping them chiral. A closed component of the dividing curve that bounds a disc detects the existence of such a Skyrmion---this is Giroux's criterion~\cite{geiges2008}. The cholesteric ground state is tight, while all textures shown in Fig.~\ref{fig:euler_class} are overtwisted, as is readily seen from the dividing curve. We do not make further use of this distinction, but some of its applications are discussed elsewhere~\cite{machon2017,hu2021,han2022,pollard2023}.

\subsection{The Characteristic Foliation in the Presence of Defects}
\label{subsec:local_obstructions2}

The arguments in the previous subsection go through so long as the surface $S$ does not intersect defects in the director. If there are defects, then the characteristic foliation will also have singularities whenever $S$ intersects those defects, and the behaviour at such points is not covered by the standard contact topology constructions we presented in the previous section. In Refs.~\cite{pollard2019,pollard2023} we defined a chiral defect to be a defect where the twist takes a single sign in a neighbourhood of the defect. This requirement results in a nontrivial constraint on the local structure of both point~\cite{pollard2019} and line~\cite{pollard2023} defects in cholesterics. We review briefly that part of the material in Refs.~\cite{pollard2019,pollard2023} that we make use of here. 

\begin{figure*}[t]
 \centering
 \includegraphics[width=\linewidth]{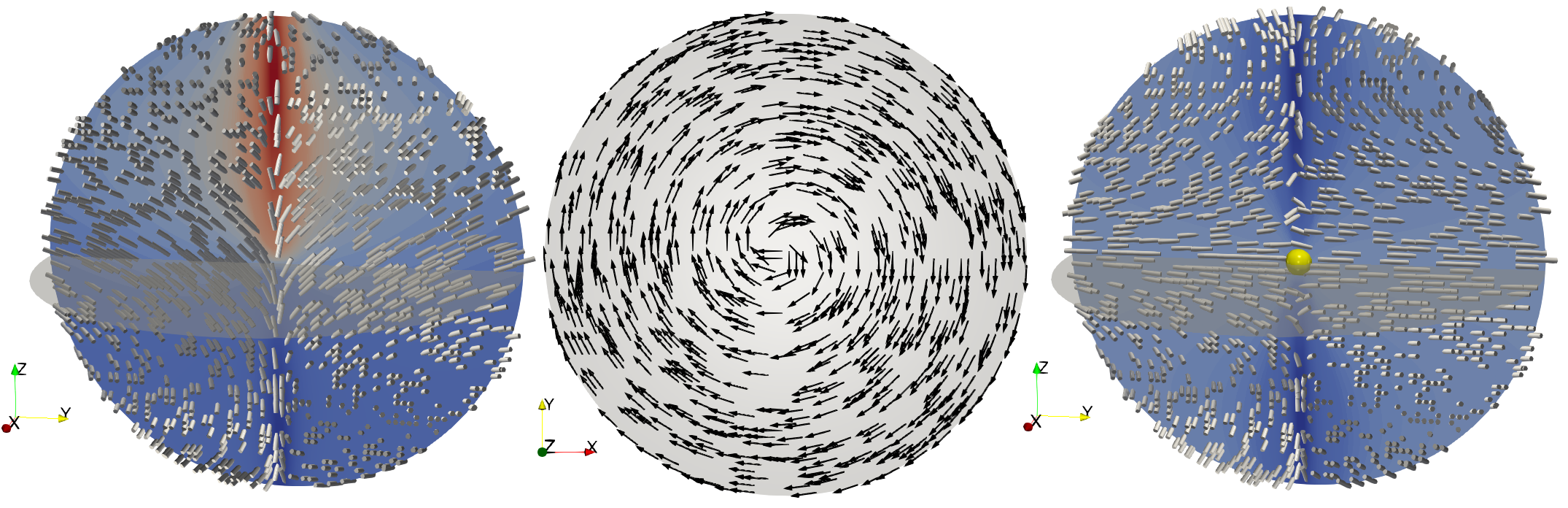}
 \caption{A characteristic foliation with a centre-type singularity is incompatible with a defect-free, chiral director. The middle panel shows a characteristic foliation (black) on the surface $z=0$. Two directors imparting this same characteristic foliation on $z=0$ are shown to the left and to the right, with the surface $z=0$ highlighted in grey. Left: the director is aligned with ${\bf e}_z$ along $r=0$, resulting a defect-free director (white) which is not chiral. The twist is shown on the slice $x=0$: blue is negative, energetically preferred, while red is positive---the twist changes sign as we pass over $z=0$. Right: the director is aligned with $z{\bf e}_z$ along $r=0$, which results in a point defect at the origin (yellow) whose structure is exactly (\ref{eq:chiral_point_defect}). However, in this case it is possible for the director to be chiral throughout.}
 \label{fig:char_fol_centre}
\end{figure*}

A generic chiral point defect is given locally by the director ${\bf n} = \pm {\bf m}/|{\bf m}|$, where
\begin{equation} \label{eq:chiral_point_defect}
    {\bf m} = (x-qyz){\bf e}_x + (y+qxz){\bf e}_y -2z {\bf e}_z,
\end{equation}
in local coordinates~\cite{pollard2019}. The projection of the director into the surface $z=0$ has a purely radial singularity, and so the characteristic foliation has a centre, the divergence of which vanishes. Equivalently, along a surface of constant $z$ $n_\perp = c_1\zeta$ for $c_1=(1+qzi)$, which is purely real at $z=0$. Combined with the arguments in the previous subsection, we see that the characteristic foliation on a surface may possess singular points where $\textrm{div}\, {\bf f}=0$ only if that point is a defect in the director; equivalently, such a point indicates either the presence of a defect of a reversal of handedness, and we cannot tell which case we are dealing with from the characteristic foliation alone. The centre-type singularity, as well as the two directors imparting it, is shown in Fig.~\ref{fig:char_fol_centre}. The leftmost panel shows the case of a reversal of handedness, while the rightmost panel shows a chiral point defect. 

Radial hedgehog point defects are incompatible with a constant sense of handedness---indeed, a chiral director field may never be transverse to a sphere~\cite{confoliations,pollard2019}. To see this, let $S$ be any surface to which the director is transverse, pointing out, say. The director ${\bf n}$ provides a local trivialisation of a neighbourhood of the surface, making it homeomorphic to $S \times [0,1]$, where the $[0,1]$ factor represents a distance outward from $S$ along the integral curves of ${\bf n}$. It is convenient to consider the 1-form $\eta$ dual to the director, which can then be interpreted as a vector-valued connection 1-form on $S$ that associates to a vector field tangent to $S$ its component along ${\bf n}$. This provides us with a geometric way of measuring twisting near $S$. Informally, consider a disc $D$ on the surface $S$, with boundary a closed curve $\gamma$, and pick a point $p$ on $\gamma$. Starting from $p$, we may lift the curve $\gamma$ off the surface so that it is everywhere orthogonal to the director ${\bf n}$. This results in a new curve $\gamma^\prime$ which will not in general close up, and the displacement between the endpoints of $\gamma^\prime$ gives a measure of the handedness of the director at the point $p$.

Now we give a more formal description. By construction the characteristic foliation is a vector field that is everywhere orthogonal to the director, and away from any singularities it defines for us an orthonormal frame ${\bf e}_1 = {\bf f}/|{\bf f}|$, ${\bf e}_2 = {\bf n} \times {\bf e}_1$, ${\bf e}_3 = {\bf n}$. The gradients of the director may be expressed in terms of the Lie brackets between the components of the frame~\cite{confoliations,pollard2021,dasilva2021}, and in particular the twist is ${\bf n} \cdot \nabla \times {\bf n} = - {\bf n} \cdot [{\bf e}_1, {\bf e}_2]$. If $\gamma$ is a closed integral curve of ${\bf f}$, then the displacement between the endpoints of the lifted curve $\gamma^\prime$ at the point $p$ is exactly the value of ${\bf n} \cdot [{\bf e}_1, {\bf e}_2]$ at $p$. We define a vector-valued 2-form $\Phi$ on $S$, taking values in the normal direction, by setting $\Phi({\bf u}, {\bf v}) = \left({\bf n} \cdot [{\bf u}, {\bf v}] \right) {\bf n} $ for any pair of vector fields ${\bf u}, {\bf v}$ on $S$. Locally this is the curvature form of the connection 1-form $\eta$, so that on any disc $D$, $\Phi = d\eta$~\cite{note1}. When $D$ has boundary curve $\gamma$, Stokes' Theorem gives an equality
\begin{equation}
    \int_D \Phi = \int_\gamma \eta.
\end{equation}

Now we demonstrate that a chiral director may not be transverse to a sphere $S$. Decompose $S$ into two discs $D_1, D_2$, oriented so that the normal to each disc points out of the sphere, with common boundary a closed curve $\gamma$. If the director is right-handed near $D_1$, the displacement of the endpoints of $\gamma$ is positive, and then we must have
\begin{equation}
    0 < \int_{D_1} \Phi = \int_\gamma \eta,
\end{equation}
when $\gamma$ is oriented as the boundary of $D_1$. Since $\gamma$ is also the boundary of $D_2$ but with the opposite orientation, then when ${\bf n}$ is also right-handed near $D_2$ we have,
\begin{equation}
   0 < \int_{D_2} \Phi = -\int_\gamma \eta.
\end{equation}
Thus we arrive at a contradiction---if ${\bf n}$ is transverse to a sphere, it cannot be chiral. This shows no hedgehog defect may be chiral. A further consequence of this is that radial boundary conditions on a spherical droplet of cholesteric imply the presence of regions where the twist of the director has sign opposite to that which is energetically preferred, and these regions are topologically protected so long as the anchoring is sufficiently strong~\cite{pollard2019}. 

Now we turn to singular lines. The classification of chiral line singularities is more involved than the classification in an achiral nematic~\cite{pollard2023}, making use of dividing curves and other methods of contact topology to analyse the local structure on a torus around the singular line. In Ref.~\cite{pollard2023} we focused on the case of disclination lines, but the classification scheme derived there applies equally to removable singular lines with integer winding. 
%
%While all chiral point defects are locally overtwisted, as can be seen from the dividing curve on a convex sphere around the defect~\cite{pollard2023}, chiral line singularities may be either tight or overtwisted in a local neighbourhood of the line. We only discuss the tight class here. 
Chiral line singularities may be either tight or overtwisted---the fundamental dichotomy of contact structures~\cite{geiges2008,machon2017,pollard2023}---in a local neighbourhood of the line; we only discuss the tight class here. Tight singular lines have the property that, after perhaps making a small deformation of the director that preserves the chirality, the winding of the director around the boundary of a disc orthogonal to the director is a constant that is independent of the choice of disc; that is, there is no variation in the winding of the profile. The classification then further splits into two classes, the $\tau$ and the $\chi$ lines~\cite{kleman1969, bouligand1978, pollard2023}, according to whether or not the line is also a singularity of the cholesteric pitch axis. 

Just as we may give a normal form for a chiral point defect, Eq.~(\ref{eq:chiral_point_defect}), we may give normal forms for the director in a neighbourhood of each class of singular line. We write these in terms of a local coordinate $z$ along the singular line and coordinates $x,y$ in the planes orthogonal to the singular line. Any tight singular line has a well-defined winding number $k$, a half integer, with fractional $k$ corresponding to a disclination line and integer $k$ corresponding to a removable singularity. 

The familiar form of a $\tau$ line is given by taking the pitch axis along a vector field ${\bf m}_k$ in the $\{{\bf e}_x,{\bf e}_y\}$-plane with winding number $k$ (any half-integer) and setting
\begin{equation}
    {\bf n}^{\tau}_{k} = \sin\psi \,{\bf e}_z + \cos\psi \,{\bf m}_{k} \times {\bf e}_z ,
    \label{eq:tight_model_tau}
\end{equation}
where $\psi$ is a helical phase increasing along the pitch axis. When $k=1$, this yields the radial and rotational singularities considered in Section \ref{sec:achiral_escape}, depending on whether we choose the axis ${\bf m}_1$ to be aligned with the radial or rotational direction. We emphasise that these two cases are topologically equivalent to one another as chiral directors via a free homotopy preserving the sense of handedness~\cite{alexander2012}, but are distinct when there are boundary conditions.

Cholesterics also exhibit a different type of singularity, the $\chi$-line, which is singular for the director but not the pitch axis. All $\chi$-lines have a local neighbourhood equivalent to a director of the form
\begin{equation} \label{eq:chi_lines}
    {\bf n}^\chi_{k,q} = \cos(k\theta + qz){\bf e}_x + \sin(k\theta + qz) {\bf e}_y,
\end{equation}
where $k$ is again a half integer constant that describes the winding of the profile and $q$ is a half integer that gives the number of full $2\pi$ rotations of the profile as we move along the line---it is the Thurston--Bennequin number of a Legendrian curve parallel to the singularity. Neither $k$ nor $q$ is a topological invariant in an achiral nematic, but both must be conserved in a cholesteric~\cite{pollard2023} (if a uniform handedness is maintained). The description of integer-winding $\chi$-lines is further simplified by employing the complex notation introduced in Eq.~(\ref{eq:char_fol_chiral1}). Any complex function gives rise to an associated real vector field on the plane by taking the $x$ and $y$ components of the vector field to be the real and imaginary parts of this function. In this notation a $\chi^{k}$-line, $k > 0$, corresponds to the function
\begin{equation} \label{eq:chi_lines_complex}
    m = \mathrm{e}^{i\alpha} \,\mathrm{e}^{iqz} \,\zeta^k ,
\end{equation}
where $\alpha$ is a constant phase. To describe a $\chi^{-k}$-line, we replace $\zeta^k$ with the complex conjugate $\bar{\zeta}^{k}$. For a $\chi^{+1}$-line, the projection into a $z$-plane has $c_1 = \mathrm{e}^{i(\alpha+qz)}$ in (\ref{eq:char_fol_chiral1}), and hence the sign of its imaginary component varies with $z$ when $q \neq 0$. This is key to understanding the possibilities for the chiral escape of the lines, which we discuss in \S\ref{sec:chi1}.

\section{Non-Abelian Chiral Escape and Toroidal Twist Solitons}
\label{sec:tau1}

Both the rotational and radial forms of the $\tau^{+1}$-line (\ref{eq:tight_model_tau}) can be removed via escape while keeping the director chiral, as we already described in Section \ref{sec:achiral_escape}. For tangential boundary conditions the direction of this escape is enforced by the handedness of the material, Fig.~\ref{fig:escape}(b,c). However, for radial boundary conditions on the cylinder the two directions of escape are equivalent and have equal energies, and as such it is possible for regions of escape up to coexist with regions of escape down in the cylinder. Recall that these two cases are associated with distinct elements of $\pi_2(\mathbb{RP}^2,\mathbb{RP}^1)$, namely $(1,0)$ for escape up and $(0,-1)$ for escape down which implies that interfaces between two regions that escape in opposite directions must contain a point defect. These have to adopt a special local structure, Eq.~(\ref{eq:chiral_point_defect}), in order to be uniformly chiral~\cite{pollard2019}. 

It follows that for escape in a cholesteric the order is important, {\sl vis-\`a-vis} escape down-escape up or escape up-escape down. This gives both a curious example of non-Abelian structure and also of topological chiral frustration. This is because in one case the defect on the interface has a hyperbolic structure and can be made chiral, while in the other case it is a radial hedgehog which is fundamentally incompatible with chirality~\cite{pollard2019}.

\begin{figure*}[t]
 \centering
 \includegraphics[width=1.0\linewidth]{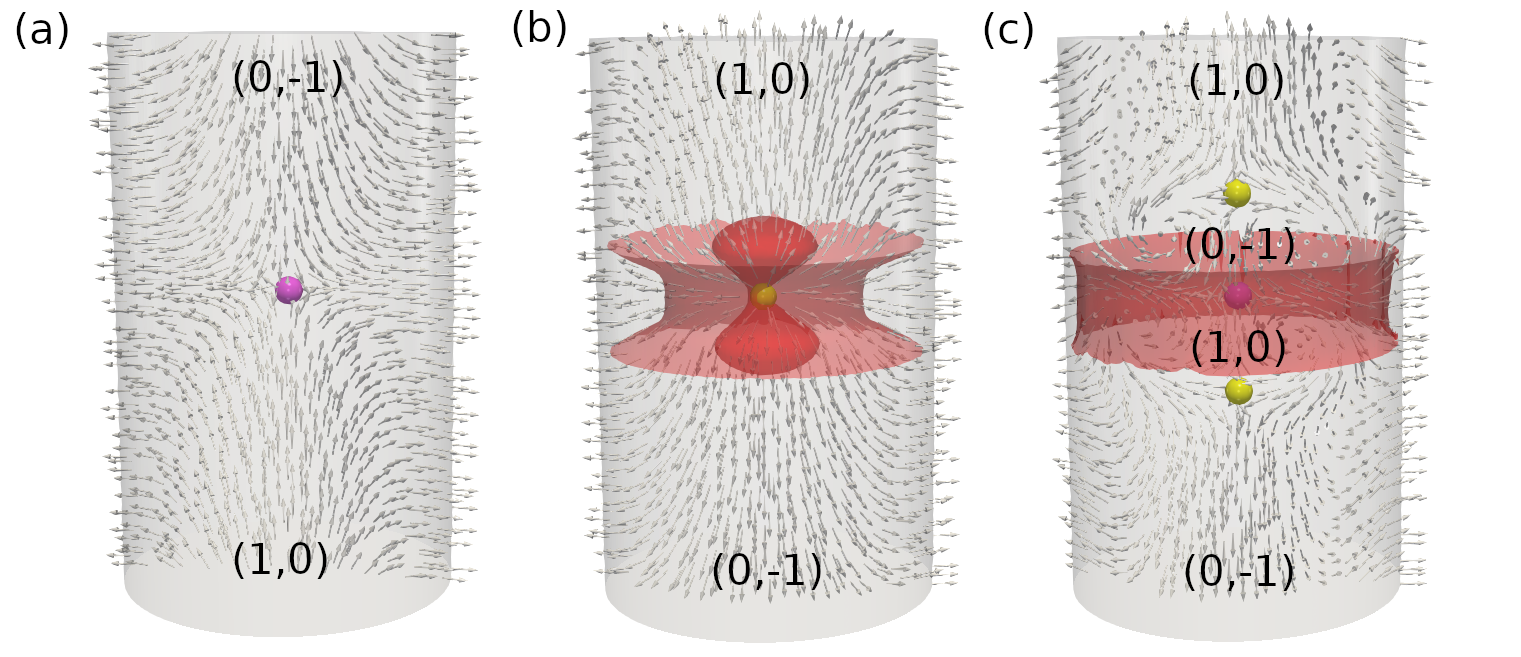}
 \caption{Escape of a cholesteric in a cylindrical capillary with radial anchoring on the boundary. We have oriented the director so that it points outwards on the boundary, a choice which also defines `up' in the cylinder. Regions of escape up and escape down have the same energy, and may coexist. On the interface between such regions there must be a defect. (a) When a region of escape down is placed above a region of escape up, the defect is hyperbolic and chiral (purple sphere). (b) In the opposite case, the defect is a radial hedgehog, fundamentally achiral. The isosurface ${\bf n} \cdot \nabla \times {\bf n} = 0$ (red) shows a twist soliton is pinned to the hedgehog, and there is also a ring atatched to the boundary. (c) Given sufficiently strong chirality, the radial defect in panel (b) converts into a string of hyperbolic defects of alternating charge $+1$ (yellow) and $-1$ (purple). A twist soliton (red) remains close to the boundary, although it shrinks in size compared to the lower-chirality case. In each case we have labelled the upper and lower boundary discs of the cylindrical section by the element of $\pi_2(\mathbb{RP}^2,\mathbb{RP}^1)$ the director corresponds to on that disc. We have also indicated this between the defects in (c)---in these regions there is an additional $\pi$ twist in the director between the core and the boundary. }
 \label{fig:cylinder_defects}
\end{figure*}

To illustrate this, we perform a numerical simulation in a cylinder of length $L$ with enforced radial boundary conditions. The initial condition is a radial director forced to escape down in the upper part and up in the lower part of the cylinder. Starting from this initial condition, we then minimise the energy~\eqref{eq:Frank_energy}. The result is shown in Fig.~\ref{fig:cylinder_defects}. The defect on the interface is a hyperbolic defect with winding $-1$---the purple sphere in the figure indicates its position---and the director is chiral throughout. The other case, escape up-escape down, is shown in Fig.~\ref{fig:cylinder_defects}(b). The simulation is performed identically, reversing the direction of forced escape in the initial condition. In this case a hedgehog defect appears on the interface, and it is pinned to a region of reversed handedness due to the constraint described---the boundary of this region is shown in red. The local structure of this defect is similar to the hedgehogs that sit near the boundary of cholesteric droplets with radial anchoring~\cite{posnjak2017,pollard2019} and to the defects in the `twiston'~\cite{ackerman2016}. 

The situation shown in Fig.~\ref{fig:cylinder_defects}(b) is stable at low chirality, $2\pi/q_0 \sim L/2$, but in a simulation performed at higher chirality, $2\pi/q_0 \sim L$, the director removes the twist soliton attached to the defect by reconfiguring the hedgehog into a hyperbolic defect with charge $-1$ (magenta), spawning two additional charge $+1$ hyperbolic defects (yellow) to maintain the overall charge. This is shown in Fig.~\ref{fig:cylinder_defects}(c). The director is now uniformly right-handed around this trio of defects. In both the low and high chirality cases there is an additional twist soliton pinned to the boundary, which takes the form of a thin ring. This twist soliton is topologically protected and also arises from the constraint described in Section~\ref{subsec:local_obstructions2}---there is a sphere to which the director is transverse, and even after expanding the interior defect this sphere remains. In a neighbourhood of this sphere it is not possible to be chiral, and hence there is a twist soliton which will be energetically stable given sufficiently strong surface anchoring. A similar structure is observed in spherical droplets with radial anchoring~\cite{pollard2019}---the local structure around the twist soliton is identical to that described in Ref.~\cite{pollard2019}. 

\section{Frustration of Escape For Chi-Lines and Spherical Twist Solitons}
\label{sec:chi1}

We now examine the possibility of removing $\chi^{+1}$ lines by escape. Setting $k=+1$ in~\eqref{eq:chi_lines} gives the local model
\begin{equation}  \label{eq:chi1}
    {\bf n}^\chi_{1,q} = \cos qz \ {\bf e}_r + \sin qz \ {\bf e}_\theta,
\end{equation}
Equivalently, we take $k=+1$ in the complex form~\eqref{eq:chi_lines_complex}. The projection of the director into a surface of constant $z$ is then given by~\eqref{eq:char_fol_chiral1}: for a $\chi^{+1}$ line we have $c_1 = |c_1| \,\mathrm{exp}(i q z)$. The constraint expressed in~\eqref{eq:char_fol_chiral2} applies to attempts to escape these singular lines: $\mathrm{Im}\,c_1$ must alternate in sign along $z$, and therefore the line cannot have both a uniform handedness and constant direction of escape.

If the director escapes up everywhere then the twist must change sign over each surface $\mathrm{Im}\,c_1 = 0$, and there will be twist solitons in the regions where $\mathrm{Im}\,c_1 > 0$. Close to the changes in handedness, the director is as shown in the leftmost panel of Fig.~\ref{fig:char_fol_centre}. Alternatively the director may escape up in regions where $\mathrm{Im}\,c_1 < 0$ and down in regions where $\mathrm{Im}\,c_1 > 0$. In this case there will be point defects at the origin for each value of $z$ where $\mathrm{Im}\,c_1 = 0$. The local structure of the these points defects is Eq.~(\ref{eq:chiral_point_defect}), as shown in the rightmost panel of Fig.~\ref{fig:char_fol_centre}.

\begin{figure*}[t]
 \centering
 \includegraphics[width=1.0\linewidth]{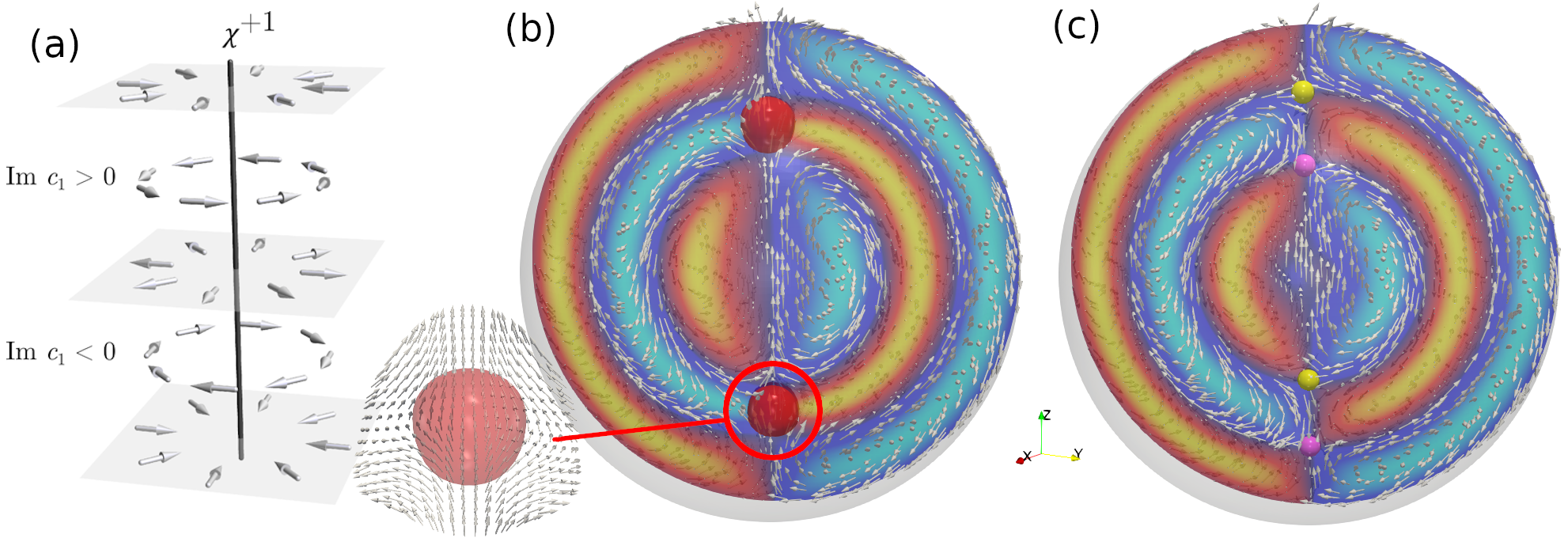}
 \caption{Escape of singular lines with $+1$ winding in a cholesteric. (a) Schematic of a $\chi^{+1}$ line. (b,c) Escape of $\chi^{+1}$ line aligned with the $z$-axis in a spherical droplet. If the director escapes up everywhere along the $z$-axis, (b), then twist solitons (red) are forced to be present and are topologically protected. The energetically-preferred option is for the director to escape alternately up and down, (c), resulting in a string of defects of alternating topological charge (magenta and yellow spheres). }
 \label{fig:chi1}
\end{figure*}

The $\chi^{+1}$ line arises naturally in a spherical geometry, which gives rise to texture with cholesteric pseudolayers that consist of concentric spheres. The string of defects that results from removing this singular line is a well-known texture occurring in spherical droplets with tangential anchoring~\cite{sec2012}, as well as in spherical shells~\cite{darmon2016, darmon2016b}. In Fig.~\ref{fig:chi1}(b,c) we show textures in a cholesteric droplet with degenerate planar anchoring on the boundary. The initial director is tangent to concentric spheres and rotates between the $\theta$ and $\phi$ directions as we move radially, where $r,\theta,\phi$ denote the usual spherical coordinates. Close to the singular line, which falls along the $z$-axis, the director is modelled on Eq.~(\ref{eq:chi1}). 

In Fig.~\ref{fig:chi1}(b) we show the result of escaping upwards everywhere along the line. To simulate this texture we force the director to escape-up along the $z$-axis and then allow it to relax to a metastable state by minimising the energy~\eqref{eq:Frank_energy}. The resulting director (grey arrows) is shown on a slice across the droplet, which is coloured according to whether or not the director points into (blue) or out of (yellow) the surface. Boojums sit at the north and south pole of the sphere. The surface ${\bf n} \cdot \nabla \times {\bf n} = 0$ is shown in red. It comprises two small regions, inside which ${\bf n} \cdot \nabla \times {\bf n} > 0$ and the handedness is opposite to that preferred by the energy (with $q_0>0$). These twist solitons have a different structure from the ring described in the previous section. The local structure of the director around the twist soliton is shown in more detail in the bottom left inset. 

Initially these twist solitons stretch between the spheres whose radius $r$ is such that $\sin q r = 0$. As the director rotates by $\pi$ between these surfaces, we might expect that the size of the twist soliton is set to be a half-pitch length. In fact, minimising then energy causes them toy become slightly smaller than a half-pitch, but nonetheless they remain stable and do not shrink; exactly what sets the size is an intriguing open problem. In an achiral nematic with equal elastic constants the texture shown in Fig.~\ref{fig:chi1}(b) decays and converts to a state that is uniform in the bulk---we have confirmed this by minimising the energy~\eqref{eq:Frank_energy} with $q_0=0$. This establishes that it is indeed chirality and not anisotropy in the elastic constants that stabilises this texture, even though the texture itself is not uniformly chiral. 

The energetically-favoured path to removing the $\chi^{+1}$ line is to escape alternately up and down, resulting in a uniformly chiral texture containing a string of chiral point defects with alternating topological charge $\pm 1$. This case is illustrated in Fig.~\ref{fig:chi1}(c), with defects indicated by yellow (charge $+1$) and magenta (charge $-1$) spheres. This results from numerically minimising~\eqref{eq:Frank_energy} when the director is initialised with a singular $\chi^{+1}$-line along the $z$-axis, whereas to simulate Fig.~\ref{fig:chi1}(b) we must force the director to escape up before allowing it to relax. We emphasise however that both of the textures shown in Fig.~\ref{fig:chi1}(b,c) are numerically stable across the length of our simulation, suggesting that both textures can be realised in experiments if the direction of escape is controlled, for example with an applied field. Strings of defects with a very similar local structure have also been observed in experiments on chromonic liquid crystals confined in cylindrical capillaries~\cite{eun2021}. The point defects move very close together, with the distance between them being significantly smaller than one half-pitch, however they do not annihilate---as with the twist solitons, it is not immediately clear what sets this lengthscale, and understanding why these structures are stable remains an open problem. 

The frustration described here does not arise for a $\chi^{-1}$ line, as can be readily checked using the calculations of Section \ref{sec:local_obstructions1}: for a $\chi^{-1}$ line we take $c_{-1} = |c_{-1}| \,\mathrm{exp}(i q z)$ and $c_1$ independent of $z$. Such lines can always escape to produce a nonsingular texture---whether or not they do is then a question of energetics. Negative winding $\chi$-lines are not naturally associated with traditional geometries like spheres and cylinders, but could be stabilised in handlebody droplets, or the complement of handlebody-shaped colloidal inclusions.

\section{Splittings of Singular Lines}
\label{sec:chi2}

The frustration discussed in the previous section applies not just to $\chi^{+1}$-lines, but more generally to all $\chi$-lines with winding $k\neq -1$. The most pertinent case to consider is the $\chi^{+2}$-line, as these arise naturally in a variety of experimental settings: they have been observed in the Smectic A-Cholesteric phase transition~\cite{cladis1979}, and also arise naturally in a spherical geometry with tangential boundary condition\cite{sec2012}, with the director being tangent to concentric spheres. In the latter geometry the $\chi^{+2}$-line is converted into a $\lambda^{+1}$-line taking the form of a twisted helix. This structure is known as the `spherulitic' or `Frank--Pryce' structure~\cite{robinson1958,sec2012}. A similar escaped structure has also been observed in spherical shells~\cite{darmon2016,darmon2016b}, this time with two intertwined $\lambda^{+1}$-lines stretching between the boundaries of the shell. A similar linked pair forms the core of the heliknoton structure recently observed in cholesterics\cite{tai2019,wu2022} and predicted in magnetic systems~\cite{smalyukh2020}. The existence of these knotted structures is a direct consequence of `escape into the third dimension' being frustrated in a cholesteric. In this section we present a novel calculation describing how this frustration is relieved. 

\begin{figure*}[t]
 \centering
 \includegraphics[width=\linewidth]{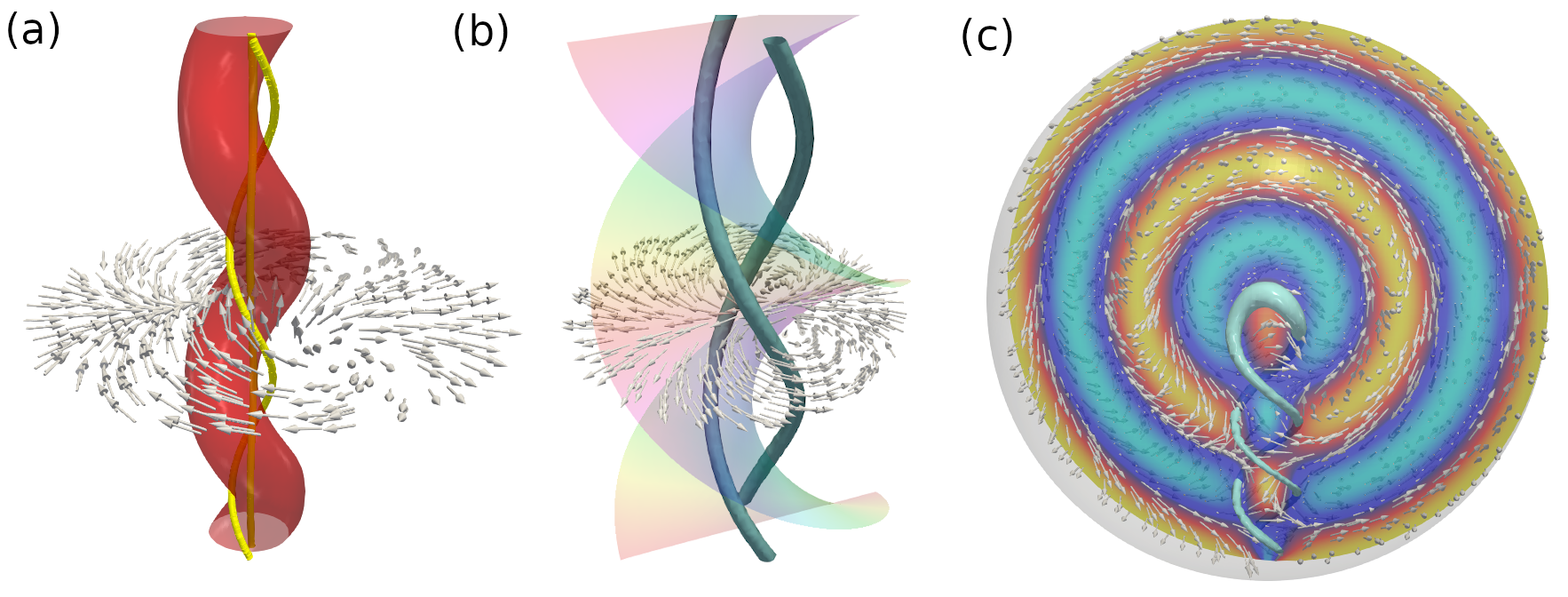}
 \caption{Formation of helical structures in cholesterics. (a) Escaping along a $\chi^{+2}$ line results in a stable region of reversed handedness (red). The director is shown on a cross-sectional disc (white arrows) The lines where the director is aligned with the ${\bf e}_z$ direction are shown as yellow tubes. The director is planar around the boundary of the illustrated disc, and so the director can be assigned an invariant $(2,0) \in \pi_2(\mathbb{RP}^2,\mathbb{RP}^1)$. (b) Instead of escaping, a $\chi^{+2}$-line in a cholesteric splits into a pair of intertwined $\tau^{+1}$-lines, which then escape into $\lambda^{+1}$-lines (pale blue isosurface). The winding of the director, shown as white sticks on a cross-sectional slice, is visualed through the Pontryagin--Thom surface where the $z$-component vanishes. (c) The $\chi^{+2}$-line arises naturally in a spherical droplet, where its escape results in a double-helix $\lambda$-line structure. The director is shown on a cross-section, coloured blue or yellow according to whether the director points into, or out of, the slice.}
 \label{fig:chi2}
\end{figure*}

First, however, we summarise the non-chiral possibilities for removal. Analogously to the $\chi^{+1}$-line, a $\chi^{+2}$-line may simply escape into the third dimension and be replaced by either a string of point defects of alternating charge $\pm 2$---this is not stable in numerical simulations---or else by a nonsingular texture in which the twist must necessarily change sign, which can be stabilised in a numerical simulation. Initialising the director (\ref{eq:chi_lines}) with $k=2,q=1$ and then enforcing escape-up results, upon numerical minimisation of the energy, in the structure shown in Fig.~\ref{fig:chi2}(a). The director on a cross-sectional disc has winding $+2$. We show the set of points where the director is ${\bf e}_z$ as yellow tubes---there are no points where the director is aligned with $-{\bf e}_z$, and there is no linking of preimages in this texture. The region of reversed-handedness takes the form of an oscillating tube, shown in red in Fig.~\ref{fig:chi2}(a). To our knowledge this texture has never been observed in an experiment, but it could be engineered using surface-patterning techniques~\cite{modin2023} or an applied field. Since the director is planar away from the original $\chi^{+2}$-line, on any cross-sectional disc we can associate it with a homotopy class in $\pi_2(\mathbb{RP}^2,\mathbb{RP}^1)$, and this class is independent of $z$ even though the director is not. Examining the characteristic foliation, we see that there are two merons of type $(1,0)$, and so this texture belongs to the homotopy class $(2,0)$ and has Skyrmion charge $Q=1$~\cite{note2}. 

In order to produce a defect-free, chiral director a $\chi^{+2}$-line must pull apart into singular lines of lower winding which can then escape while keeping the director chiral. This is exactly what is observed in the Frank--Pryce texture. The description of the splitting of integer winding line defects, their `unfoldings'~\cite{arnold, pollard2019}, is best expressed using the complex notation introduced in Eq.~(\ref{eq:char_fol_chiral1}). The unfolding of a $\chi^{+2}$-line into a pair of $\chi^{+1}$-lines is described by the complex function
\begin{equation} \label{eq:perturbed_chi2}
    m = \mathrm{e}^{i\alpha} \,\mathrm{e}^{iqz} \bigl( \zeta - R \,\mathrm{e}^{iq^{\prime}z} \bigr) \bigl( \zeta + R \,\mathrm{e}^{iq^{\prime}z} \bigr), 
\end{equation}
where $R$ sets the distance between the two singularities and $q^\prime$ is a half-integer parameter that controls their rotation as we move along the $z$ axis. The two singular lines are at $\zeta = \pm R \,\mathrm{e}^{iq^{\prime}z}$. Near $\zeta = R \,\mathrm{e}^{iq^{\prime}z}$ we have 
\begin{equation}
    m = \mathrm{e}^{i\alpha} \bigl( \zeta - R \,\mathrm{e}^{iq^{\prime}z} \bigr) \,2R\,\mathrm{e}^{i(q+q^{\prime})z} + \cdots ,
\end{equation}
while near $\zeta = -R \,\mathrm{e}^{iq^{\prime}z}$ we have 
\begin{equation}
    m = \mathrm{e}^{i\alpha} \bigl( \zeta + R \,\mathrm{e}^{iq^{\prime}z} \bigr) \,(-2R)\,\mathrm{e}^{i(q+q^{\prime})z} + \cdots = \mathrm{e}^{i(\alpha+\pi)} \bigl( \zeta + R \,\mathrm{e}^{iq^{\prime}z} \bigr) \,2R\,\mathrm{e}^{i(q+q^{\prime})z} + \cdots .
\end{equation}
Since the $\zeta$ coordinate system does not depend on $z$, the rotation of the profile along each of these singular lines is determined entirely by the half-integer $q+q^\prime$. When $q^\prime$ is an integer, the $\chi^{+2}$-line pulls apart into a pair of $\chi^{+1}$-lines that are interlinked with one another $q^\prime$ times. Relative to the original line, these two lines are helices which are left-handed for $q^\prime < 0$ and right-handed for $q^\prime > 0$; they are parallel for $q^\prime = 0$. The profile winds through $q+q^\prime$ full turns as we move along the line---that is, the director around each of these lines is locally modelled on ${\bf n}^\chi_{1, q+q^\prime}$. In particular, by choosing $q+q^{\prime} = 0$ the profile along each singularity is constant and they have the form of a $\tau^{+1}$-line. 

This splitting has an interpretation in terms of the C\u{a}lug\u{a}reanu theorem~\cite{calugareanu1959,calugareanu1961,white1969,fuller1971}. Recall that for a closed curve $K$ with any given framing we define the topological self-linking number $\textrm{SL}(K)$ to be the linking number between $K$ and the give $K^\prime$ obtained by displacing $K$ along its framing. We also define the geometric twist $\textrm{Tw}(K)$ and writhe $\textrm{Wr}(K)$ of the curve. Informally, the twist measures the angular rotation of the frame as we move along the line, and the writhe measures the `coiling' of the curve in three-dimensions. Together these three quantities satisfy the relationship
\begin{equation} \label{eq:calugareanu}
    \textrm{SL}(K) = \textrm{Tw}(K) + \textrm{Wr}(K).
\end{equation}
A singular line in a chiral material comes with a natural framing, the `contact framing'~\cite{geiges2008,pollard2023}. The singular line $K$ with local neighbourhood described by the idealised director~\eqref{eq:chi_lines} and the contact framing has self-linking number $q$, twist $\textrm{Tw}(K) = q$, and vanishing writhe, which evidently satisfy~\eqref{eq:calugareanu}. If the curve deforms via an isotopy the self linking number is always fixed, but the C\u{a}lug\u{a}reanu theorem shows that we may exchange twist for writhe. The splitting of a $\chi^{+2}$-line presents an analogous situation: when $q^\prime=-q$, the $\tau^{+1}$-lines that result from splitting the $\chi^{+2}$-line have a contact framing with zero self-linking number, $\textrm{Tw}(K)=q$ and $\textrm{Wr}(K) = -q$, so we see that the splitting can be seen, loosely, as a conversion of self-linking into writhe.

We can escape along the resulting $\tau^{+1}$-lines and remove them while keeping the director chiral if and only if the profile is constant---i.e. when $q^\prime=-q$---as this ensures a constant sign of the term $\mathrm{Im}\,c_1$ appearing in (\ref{eq:char_fol_chiral2}). The direction of the escape is controlled by the sign of this term. Consider the value of $\mathrm{Im}\,c_1$ on the surface $z=0$. The two zeros of $m$ are singularities of the characteristic foliation with local profile equivalent to $2R i \,\mathrm{e}^{i\alpha} \zeta$ and $2R i \,\mathrm{e}^{i(\alpha+\pi)} \zeta$, respectively, and at each point we have
\begin{align}
    & \Bigl. \mathrm{Im}\,c_1 \Bigr|_{\zeta=R} = 4R \sin\alpha , 
    & \Bigl. \mathrm{Im}\,c_1 \Bigr|_{\zeta=-R} = 4R \sin(\alpha+\pi) = -4R \sin\alpha .
\end{align}
Thus, one line escapes up and the other escapes down. If both escape up, or both down, then the escaped texture does not have a consistent handedness. 

A simulation of the resulting structure is shown in Fig.~\ref{fig:chi2}(b). The $\lambda^{+1}$-lines that result from the escape can be visualised via the tensor $\Delta$, the anisotropic part of the director gradients~\cite{machon2016,pollard2021,dasilva2021}---these are the pale blue tubes in Fig.~\ref{fig:chi2}(b). The coloured surface is the Pontryagin--Thom surface~\cite{chen2013} where the $z$ component of the director vanishes, coloured according to the angle between the $x$ and $y$ components. This clearly shows the screw-like structure of the texture. The director is illustrated on a cross section, where the local structure of an unfolded $+2$ singularity is clear. This director corresponds to the element $(1,-1) \in \pi_2(\mathbb{RP}^2,\mathbb{RP}^1)$, and it therefore has vanishing Skyrmion charge while also not being equivalent to a uniform director. 

A $\chi^{+2}$-line arising in a spherical geometry is removed by exactly the process we have just described, resulting in the spherulitic texture shown in Fig.~\ref{fig:chi2}(c). We again visualise the structure through a level set of $\Delta$. We simulate this by initialising the director to be tangent to concentric and have a $+2$-winding singular line along the negative $z$-axis. The helical structure appears spontaneously upon numerical minimisation of~\eqref{eq:Frank_energy}. By contrast with the $\chi^{+1}$, the escaped director is nonsingular and chiral throughout the droplet.

\begin{figure*}[t]
 \centering
 \includegraphics[width=\linewidth]{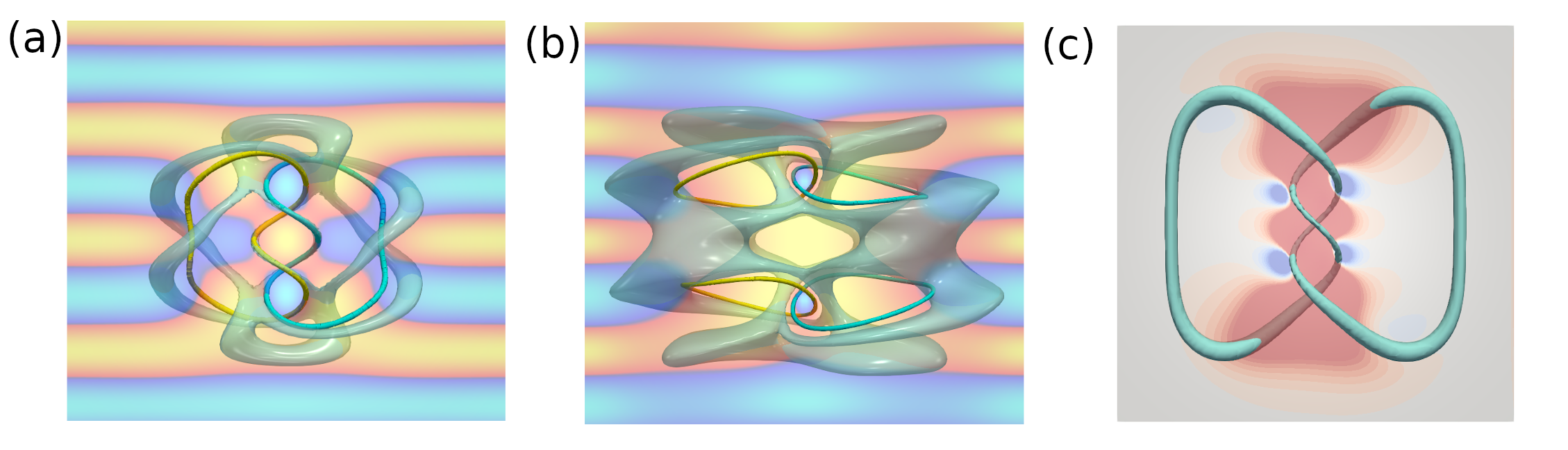}
 \caption{ (a) An escaped structure lies at the heart of a heliknoton, where the linking of the $\lambda$-lines (pale blue isosurface) results in a nontrivial Hopf invariant, as described in the text. (b) The structure shown in (a) reconfigures itself via a homotopy so that the preimages of the director (yellow, blue tubes) form a series of $q$ Hopf links, which is the configuration observed in experiments~\cite{tai2019}. (c) The model Eq.~(\ref{eq:perturbed_chi2}) can also be used to describe a single knotted $\lambda$-line, for example a trefoil knot, by choosing a non-integer $q^\prime$. Although it contains a knotted soliton, this texture has vanishing Hopf invariant, and the director points along ${\bf e}_z$ at every point along the $\lambda$-line. Moreover, this texture cannot be chiral and is numerically unstable in a cholesteric.}
 \label{fig:heliknotons}
\end{figure*}

Another important example of a texture resulting from an escaped $\chi^{+2}$-line is the `heliknoton'~\cite{tai2019,smalyukh2020,wu2022}. We show a simulation in Fig.~\ref{fig:heliknotons}(a,b). We continue to visualise the $\lambda$-lines through a level surface of $\Delta$~\cite{machon2016} (pale blue tubes) and also show a pair of preimages, curves where the director is aligned with ${\bf e}_z$ (yellow) and $-{\bf e}_z$ (blue). We show a cross-sectional convex surface with the dividing curve in black, coloured according to whether the director points into (blue) or out of (yellow) the surface---this clearly shows the disruption to the cholesteric layer structure resulting from the heliknoton. We simulate the heliknoton inside a cube with periodic boundaries, with an initial condition constructed as follows. We take the background to be the standard cholesteric state, and then remove a ball from the interior of the simulation box and replace the director inside that ball with the $\chi^{+2}$-line given by Eq.~(\ref{eq:chi_lines}). We make no attempt to match this onto the background director. Upon numerical minimisation of the energy the discontinuity over the boundary of the ball is removed and the $\chi^{+2}$-line escapes exactly as described above, resulting a knotted helix appearing in both the $\lambda$-lines and preimages. In order to match on to the background, these lines continue around the boundary of the sphere, creating a pair of linked curves. When the core is modelled on ${\bf n}^\chi_{+2, q}$, these two curves are linked $q$ times---in Fig.~\ref{fig:heliknotons}(a,b) we show $q=2$. Continuing to minimise the energy results in the director contracting to split the linked preimages into a series of $q$ separate Hopf links, Fig.~\ref{fig:heliknotons}(b), while the $\lambda$-lines remain a linked helix. This process occurs via a chiral homotopy, and as such does not create any singularities in the director or regions of reversed handedness. At its core each of these Hopf links still has the same structure of an escaped $\chi^{+2}$-line of type ${\bf n}^\chi_{+2, 1}$. This is the heliknoton observed by Tai and Smalyukh~\cite{tai2019}. An escaped ${\bf n}^\chi_{+2, 1}$ line also forms the central part of the Hopfions produced by Ackerman and Smalyukh in an earlier experimental study~\cite{ackerman2017}.  

A similar construction leads to knotted $\lambda$-lines, for example the trefoil shown in Fig.~\ref{fig:heliknotons}(c) which is realised by $q=0, q^\prime=-3/2$ in Eq.~(\ref{eq:perturbed_chi2}). We again simulate this inside a cube with periodic boundaries, with an initial condition constructed as described for the heliknoton. In order to be defect free the director has to escape the same direction everywhere along the line, and hence it must escape along $+{\bf e}_z$ everywhere in the core region. It follows that the director cannot be chiral---indeed, the twist on a slice shows regions of both left- and right-handedness. The director in this core area is similar to that shown in Fig.~\ref{fig:chi2}(a), with a helical twist soliton. This structure is not stable in a cholesteric. Knotted $\lambda$-lines in the form of a trefoil can be stabilised in a cholesteric, but the structure is quite different from that shown in Fig.~\ref{fig:heliknotons}(c).

Similar calculations describe the splitting of any $\chi^{k}$-line. To our knowledge such lines have not been observed in experiments except for $k = +1, +2$. Negative winding lines could arise naturally inside handlebody droplets~\cite{pairam2013} or on the complement of handlebody-shaped colloidal particles~\cite{smalyukh2018}, and it may be possible to generate a general $\chi^{k}$-line using surface-patterning techniques~\cite{modin2023}. We have performed numerical simulations of a $\chi^{-2}$-line, which result in splitting before it escapes into a pair of linked $\lambda^{-1}$-lines, completely analogously to the case of the $\chi^{+2}$-line. Such lines could therefore potentially result in complex knotted structures with nontrivial Hopf invariants, distinct from but analogous to heliknotons. This possibility remains to be explored. 

There are many different ways a singular $\chi^{k}$-line can be decomposed into singularities of a lower order, with the possibilities described by singularity theory~\cite{arnold,pollard2019}. The most simple is a symmetric unfolding into $|k|$ generic singularities of winding $\pm 1$ arranged at equidistant points on a circle, which can expressed in terms of the $|k|^\text{th}$ roots of unity. This may not be the energetically-optimal choice however. Results from contact topology~\cite{geiges2008} imply that the only constraint on making the resulting texture chiral is the local structure of the singular lines, not their number or their geometry---in theory it is possible to realise any braid, and the only constraints come from energetics. 

\begin{figure*}[t]
 \centering
 \includegraphics[width=0.25\linewidth]{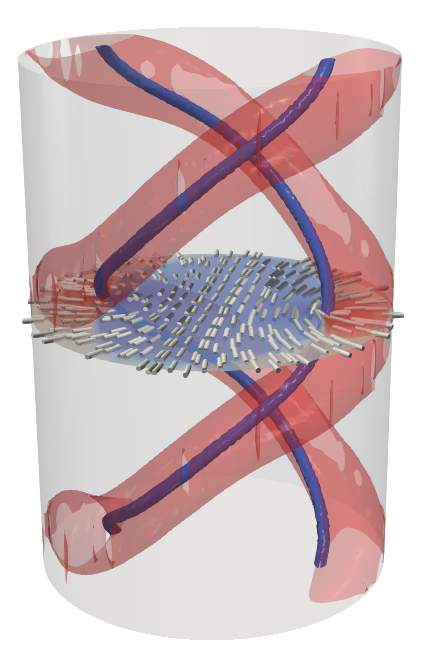}
 \caption{Rather than escaping, a $+1$ singular line in a cylindrical capillary may pull apart into a pair of disclination lines (dark blue) via the same process as a $\chi^{+2}$-line. This structure is not chiral, but rather there are regions of reversed handedness (red) pinning each disclination to the boundary. The director is shown on a cross-section.}
 \label{fig:disclinations}
\end{figure*}

We close this section by describing how a $+1$ singular line splits into a pair of disclinations, a closely related process to the splitting of a $\chi^{+2}$-line. Via a simple trick, the model (\ref{eq:perturbed_chi2}) of a $+2$ defect splitting can also be used to describe the splitting of a $+1$ line into a pair of disclinations. Suppose ${\bf v} = v_x {\bf e}_x + v_y {\bf e}_y$ is a 2D vector field with a singularity of winding $k$. We can build an associated Q-tensor by taking 
\begin{equation} \label{eq:qtensor}
    Q = \begin{bmatrix}v_x & v_y \\ v_y & -v_x\end{bmatrix}.
%    Q = \begin{bmatrix} v_x & v_y & 0 \\ v_y & -v_x & 0 \\ 0 & 0 & 0 \end{bmatrix}.
\end{equation}
Notice that the eigendirections of this tensor then have a singularity of winding $k/2$. The reader may verify that taking ${\bf v}$ to have a singularity of winding $\pm 1$, $v_x = x$ and $v_y = \pm y$, does indeed give a Q-tensor describing a $\pm 1/2$ defect. 
To obtain a three-dimensional Q-tensor with the right topological structure we can simply add an extra row and column of zeros. 
%Taking ${\bf v}$ to be the vector field derived from the complex function (\ref{eq:perturbed_chi2}) with $q=0$ and interpreting (\ref{eq:qtensor}) as defining a 3D Q-tensor by adding an extra row and column of zeros then gives the desired splitting of a $+1$ singularity into a pair of $+1/2$ disclinations that are linked $q^\prime$ times. 
Taking ${\bf v}$ to be the vector field derived from the complex function~\eqref{eq:perturbed_chi2} with $q=0$ then gives the desired splitting of a $+1$ singularity into a pair of $+1/2$ disclinations that are linked $q^\prime$ times. A simulation of this structure in a cylindrical capillary with normal anchoring is illustrated in Fig.~\ref{fig:disclinations} for $q^\prime = 1$. Larger values of $q^\prime$ add additional twists, and negative $q^\prime$ is identical except the disclinations are helices with the opposite sense of handedness. 
Unlike the defect-free escaped texture shown in Fig.~\ref{fig:escape}(a), this defective texture is not chiral, and indeed each disclination is pinned to the boundary of the cylindrical capillary by a twist soliton. This also occurs with the disclinations observed in spherical droplets of cholesteric with normal anchoring~\cite{posnjak_thesis}. The presence of this region becomes clear when we realise that a disclination line with a pure $+1/2$ profile is nothing more than a hedgehog point defect blown up into a loop, and though it is possible for a local neighbourhood of the disclination to be chiral~\cite{pollard2023}, such a removal requires a large scale global reconfiguration of the director that is frustrated by the boundary conditions.

Splittings of a radial texture into a pair of helical defect lines have been observed in lyotropic chromonic liquid crystals~\cite{jeong2015}. These materials are achiral but have a small twist elastic constant, leading to an energetic preference for twisted textures but no preference for left vs right handedness. Similar helices of disclinations have also been observed in thermotropic materials~\cite{velez2021}.

\section{Discussion}
\label{sec:discussion}

The energetic preference for maintaining a uniform sense of handedness imposes strong constraints on the textures formed in a cholesteric liquid crystal, and in particular their defects. This results in a distinct topological classification from achiral nematics, for both point defects~\cite{pollard2019} and disclination lines~\cite{pollard2023}. In this paper we have shown that chirality also places constraints on coreless defects via a topological chiral frustration. The frustration of the mechanism of `escape into the third dimension' in a cholesteric results in the formation of non-Abelian and knotted structures, as well as the twist solitons which have yet to be studied in detail. While we have focused on this mechanism in cholesteric liquid crystals, we stress that our analysis applies to all types of chiral ordered media, for example chiral ferromagnets and superfluid $^3\text{He}$. 

We have focused on the removal of singular lines with winding $+1$ and $+2$ because such lines occur naturally in cylindrical and spherical geometries and are directly relevant to existing experimental settings. Singular lines with a negative winding are associated with higher-genus handlebodies and hence energetically accessible; there is great flexibility in controlling the geometry of domains and interfaces, and exploiting surface patterning, to produce such lines. 

An additional feature of our work is the development of methods from the mathematical field of contact topology. The methods of characteristic foliations and convex surface theory are powerful tools for the analysis of all materials with nematic or vector order parameters, and especially those which are chiral. These methods can also be connected directly to quantities that are observable in experiments. 

Finally, we comment on the analogous situation in smectics, a class of material that cholesterics are often compared and contrasted with. In a smectic phase the director is normal to a family of surfaces and must have vanishing twist---as cholesterics are modelled by contact structures, so smectics are modelled by foliations~\cite{poenaru1981,chen2009,machon2019}. The equivalent to a $+1$ winding meron in a foliation is a `Reeb component'~\cite{candel2000}, but these would require compression of the smectic layers and would have infinite energy, so they cannot occur in a real material. For this reason escape of $+1$ winding singular lines is also frustrated in a smectic: in fact they do not escape, and instead form the core of focal conic domains.


\begin{thebibliography}{99}
\bibitem{note1} We remark that this relationship, $\Phi = d\eta$, need not hold over a general region, only a contractible region. Consequently, the argument given here only applies to a sphere, which is the only surface which can be decomposed into a pair of discs. A chiral director may be transverse to any handlebody, for example a torus, as explained in Ref.\cite{confoliations}.
\bibitem{note2} This appears counter-intuitive, as a texture that contains no points where the director is aligned with $-{\bf e}_z$ must have vanishing Skyrmion charge. This disparity results from considering only the local contributions from the escaped $\chi^{+2}$-line and not evaluating the Euler class on a closed surface, required by integral formulas for the Skyrmion charge. In order to extend this director to one that is uniform at infinity, we require the addition of two $-1$-winding merons. If these are also escape up then their contribution to the Skyrmion charge is $-1$, resulting in the expected result $Q=0$ from an integral over a closed surface intersecting each meron. If these additional merons are escape-down, we would instead get $Q=2$, and there would indeed be two of the familiar $(1,1)$-class Skyrmions present.
\end{thebibliography}
\end{document}